\documentclass[11pt,preprint,aps,amsmath,amssymb,prd]{revtex4-1}

\usepackage{graphicx}
\usepackage{bm}
\usepackage{color}
\usepackage{url}

\newcommand{\gmm}{}
\newcommand{\gm}{} 

\begin{document}

\def\be{\begin{equation}}
\def\ee{\end{equation}}
\def\d{\mbox{\rm d}}
\def\LCDM{$\Lambda$CDM\,}

\title{Structure formation in a Dirac-Milne universe: comparison with the standard cosmological model}

\author{Giovanni Manfredi}
\email{giovanni.manfredi@ipcms.unistra.fr}
\affiliation{Universit\'e de Strasbourg, CNRS, Institut de Physique et Chimie des Mat\'eriaux de Strasbourg, UMR 7504, F-67000 Strasbourg, France}
\author{Jean-Louis Rouet}
\affiliation{Universit\'e d'Orl\'eans, CNRS/INSU, BRGM, ISTO, UMR7327, F-45071 Orl\'eans, France}
\author{Bruce N. Miller}
\affiliation{Department of Physics and Astronomy, Texas Christian University, Fort Worth, TX 76129}
\author{Gabriel Chardin}
\email{chardin@apc.in2p3.fr}
\affiliation{Universit\'e de Paris, CNRS, Astroparticule et Cosmologie,  F-75006 Paris, France}

\date{\today}

\begin{abstract}
The presence of complex hierarchical gravitational structures is one of the main features of the observed universe. Here, structure formation is studied both for the standard ($\Lambda \rm CDM$) cosmological model and for the Dirac-Milne universe, a matter-antimatter symmetric universe that was {\gm recently} proposed as an alternative ``coasting" cosmological scenario. One-dimensional numerical simulations reveal the analogies and differences between the two models. Although structure formation is faster in the Dirac-Milne universe, both models predict that it ends shortly before the present epoch, at cosmological redshift $z \approx 3$ for the Dirac-Milne cosmology, and at $z \approx 0.5$ for the $\Lambda \rm CDM$ universe.
The present results suggest that the matter power spectrum observed by the Sloan Digital Sky Survey might be  entirely due to the nonlinear evolution of matter and antimatter domains of relatively small initial dimensions, of the order of a few tens of parsecs comoving at cosmological redshift $z =1080$.

\end{abstract}

\maketitle

\section{Introduction}\label{sec:intro}

The Standard Cosmological Model ($\Lambda \rm CDM$), which has emerged in its present form over the last two decades, is capable of accurately reproducing most cosmological observations -- among others primordial nucleosynthesis, the cosmic microwave background (CMB) radiation, baryonic acoustic oscillations (BAOs), or type-1a supernovae (SN1a) luminosity distance. Not all such measurements are equally compelling, but their combination seems to require a universe composed of much more than just ordinary (baryonic) matter. Some tensions have appeared more recently between the values of the Hubble constant $H_0$ derived from the early or late universe observations \cite{Riess2020}. These discrepancies are statistically significant ($ \approx 5\sigma$). No systematic errors have been found yet, although it is too early to ascertain whether these discrepancies are a glimpse of some ``new physics" ahead {\gm \cite{Freedman2019,Freedman2020}}.

Despite its descriptive success, the \LCDM model is clearly not a finished theory. Baryonic matter -- ordinary nuclear matter of the standard model of particle physics, which is routinely observed in particle accelerators and makes up the world around us -- constitutes today less than 5\% of the total mass-energy content in the \LCDM model, the rest being composed of cold dark matter (CDM, $\approx 25\%$) and dark energy (in the form of a cosmological constant $\Lambda$, $\approx 70\%$). This is a rather unfortunate situation, which has stimulated a lot of work, both experimental (searching experimental evidence for dark matter) and theoretical (e.g., introduction of new degrees of freedom, modification of the gravitational force, and relaxing assumptions of the standard model such as homogeneity at large scales).

Several authors have noted that our universe is very close to a ``coasting" universe, i.e., a universe that neither decelerates nor accelerates, akin to the one originally proposed by Edward Arthur Milne \cite{Milne}. A review of coasting cosmologies was published recently by Casado \cite{Casado2020}.
For example, Sarkar and coworkers  have recently argued that the present SN1a data are still unable to demonstrate convincingly the acceleration of the universe expansion rate \cite{NGS, Sarkar2019}. Further, Blanchard \textit{et al.}\ \cite{Tutu2017} have shown that cosmic acceleration by all ``local" cosmological probes (redshift $z < 3$) is not statistically compelling. Finally, when a mild evolution of the SN1a luminosity is allowed, the SN1a data of the Pantheon supernovae sample \cite{Riess2018} show that the Milne universe constitutes a fairly decent fit to the SN1a data (see for example Fig.~4 of  \cite{Riess2018}). Also, it is well known that the age of the \LCDM universe is nearly identical to the age of a Milne universe, equal to $1/H_0$.

In 2012, Benoit-Levy and Chardin \cite{Benoitlevy} proposed an alternative universe where matter and antimatter
are present in equal amounts and their gravitational interaction is repulsive, which they named the ``Dirac-Milne" (D-M)
universe to highlight its two main features, namely the presence of antimatter (hence, Dirac) and its coasting expansion law (Milne).
{\gmm
This universe, analogous in its gravitational behavior to the Dirac electron-hole system, avoids annihilation
between matter and antimatter domains after cosmological recombination. Although this scenario is clearly unconventional,
it should be noted that there is to date no direct experimental evidence on the gravitational behavior of antimatter, while several
experiments are being developed at CERN to measure the acceleration of  antihydrogen atoms ``free-floating" in the gravity
field of the Earth. The first results of the Gbar \cite{Indelicato2014}, ALPHA-g \cite{alpha-g}, and AEgIS \cite{AEGIS}
collaborations are expected within a couple of years and  deviations from perfect matter-antimatter symmetry will have
profound  consequences on current cosmological theories.

With its null total mass, the D-M universe is gravitationally empty on large scales and thus displays a coasting expansion, i.e. without acceleration nor deceleration. This behavior leads to dramatically different timescales in the early phases of the Universe. For example, the Quark-Gluon-Plasma transition lasts for about one day, instead of a few microseconds as in the Standard Model, and nucleosynthesis lasts about 35 years, compared to the three minutes in the Standard Model, while recombination occurs at an age of about 14 million years,
compared to the 380 000 years of the \LCDM model.

Despite these tremendous differences in the initial timescales, the D-M universe is remarkably concordant with only one
adjustable parameter, namely $H_0$ \cite{Benoitlevy}. In particular, its age, equal to $1/H_0$, is almost identical to the age
of the \LCDM universe for $H_0 \approx 70$ km/s, the SN1a luminosity distance is very close to that of \LCDM \cite{Chodorowski_2005, Benoitlevy},
and so is its primordial nucleosynthesis \cite{Sethi_1999, Benoitlevy}. Since the distance to its horizon diverges,
it also does not suffer from the horizon problem, nor does it need primordial inflation to explain the current
homogeneity at large scales.  A more complete description of the properties of the D-M universe can be found
in Refs. \cite{Benoitlevy} and  \cite{Chardin2018}.
}

More recently \cite{Manfredi2018}, we  developed a  theoretical basis for  the D-M model, whereby the unconventional
matter-antimatter gravitational interaction can  be accounted for, at the Newtonian level, by two gravitational potentials
that obey two coupled Poisson's equations (see Table I). This model may be viewed as the Newtonian limit of some
bi-metric extension of General Relativity or, alternatively, as mentioned above, as the description of the Dirac
 ``electron-hole" system in a gravitational field.

In the same work, we provided a detailed comparison of gravitational structure formation in the D-M and Einstein-de Sitter (EdS: $\Omega_M=1, \, \Omega_\Lambda=0$) universes, using a  local 1D model embedded in a spherically expanding universe.
For both cases, we observed gravitational structure formation, with clusters and subclusters developing from an almost uniform initial condition.
Both models display power-law behavior in the wavenumber spectrum of the matter density in both the linear and nonlinear regimes. However, there is one crucial difference. Whereas for EdS the formation of structures continues indefinitely, in the following we will see that structure formation freezes out for the Dirac-Milne universe a few billion years after the Big Bang, a feature shared with the $\Lambda\rm{CDM}$ model.

In the present work, we compare the D-M universe to a universe with finite positive cosmological constant (``\LCDM"), using the same 1D Newtonian approach. It must be stressed, however, that our ``\LCDM" universe is essentially nonlinear, in contrast to the standard cosmological model, for which the evolution of the power spectrum is almost entirely linear.
This choice was dictated by our intention to closely compare how the D-M and the nonlinear ``\LCDM" models evolve from similar initial conditions at recombination.
Our numerical results reveal striking similarities, but also significant differences between the two  cosmologies, with the D-M model predictions appearing to be compatible with the hierarchical structures observed in today's universe.

A short outline of the present work is as follows: In the next section, we briefly summarize the essential futures of the D-M universe. In Sec. \ref{sec:comoving}, we define the comoving co-ordinates used in the numerical code. Some features of the co-moving equations of motion reveal an interesting relationship between the D-M and \LCDM cosmologies. The results of the computer simulations are presented in Sec. \ref{sec:numres}, which also contains a direct comparison to data from the Sloan Digital Sky Survey (SDSS). Conclusions are drawn in Sec. \ref{sec:conclusion}.

{\gmm
\section{Gravitational properties of the Dirac-Milne universe}
\label{sec:Modeldiracmilne}
}
In the Dirac-Milne universe,  whereas matter attracts matter, all other gravitational interactions are repulsive, as is summarized in Table I,  reproduced from Ref. \cite{Manfredi2018}. Hence, matter can form gravitational structures, whereas antimatter, being repelled by everything else, tends to spread across the universe. Such spread is rather uniform, but not completely: since matter repels antimatter, the latter is expelled from matter-dominated regions (galaxies) and forms a low-density almost homogeneous background distributed over the underdense regions in between matter's  structures. As in the analog electron-hole system in a semiconductor, the matter and antimatter regions are separated from each other by a depletion zone that precludes the occurrence of annihilation events, in accordance with the observations.  A simple analytic model predicts that this depletion zone occupies $\approx 50\%$ of the volume of space, as confirmed by 3D simulations that will be presented elsewhere.

\begin{table*}[]
\begin{tabular}{|c|c|c|c|}
\hline \hline
Type of matter & Type of matter & Interaction\\
\hline
\hline
$+$ & $+$ & Attraction\\
\hline
$-$ & $-$ & Repulsion \\
\hline
$-$ & $+$ & Repulsion \\
\hline
$+$ & $-$ & Repulsion \\
\hline
\end{tabular}
\caption{Interactions between matter ($+$) and antimatter ($-$) particles in the Dirac-Milne universe, from \cite{Manfredi2018}.}
\end{table*}

It must be stressed that the Dirac-Milne scenario, in the Newtonian limit, cannot be simply described by a combination of the signs of the inertial and gravitational (active and passive) masses. Instead, as shown in \cite{Manfredi2018}, the Dirac-Milne model can only be accounted for by two gravitational potentials that obey two distinct Poisson equations:
\begin{eqnarray}
\Delta\phi_{+} &=& 4\pi G (\rho_{+} - \rho_{-}), \label{poiss_diracmilne1}\\
\Delta\phi_{-} &=& 4\pi G (-\rho_{+} - \rho_{-}) \label{poiss_diracmilne2} .
\end{eqnarray}
The above model may be seen as the Newtonian limit of a bimetric gravity model.

In the forthcoming numerical simulations, we will  make the further simplifying hypothesis that antimatter constitutes a low-density background uniformly distributed everywhere in space. This approximation appears to be justified for the study of gravitational structure formation, as overdense regions are very much dominated by matter anyway. Using this approximation, one can neglect the evolution of antimatter and replace it with a homogeneous background that decreases as the inverse of the cube of the scale factor $a(t)$:
\[
\rho_{-} (\mathbf{r},t)= \rho_0/a^3 ,
\]
where $\rho_0$ denotes today's matter density. In the following, the subscript ``0" will be used systematically to refer to quantities evaluated at the present time.
Then, the Poisson equation for matter becomes
\be
\Delta\phi= 4\pi G \left(\rho - \rho_{0} a^{-3} \right), \label{poiss_diracmilne3}
\ee
where we have dropped for simplicity the subscript ``+" denoting matter.
The dilute repulsive background could be viewed as a cosmological constant that decreases with time, the corresponding vacuum energy decreasing as $a^{-3}$.
This similarity with \LCDM underpins the observation, further described in Sec. \ref{sec:numres}, that structure formation stops around the same epoch in numerical simulations of both universes.

We further note that the standard cosmological Poisson equation is:
\be
\Delta \phi= 4\pi G ( \rho- \bar{\rho})= 4\pi G (\rho- \rho_0 a^{-3}),
\ee
where $\bar\rho$ is the average matter density,
so that the D-M and $\Lambda$CDM models have a similar form for the Poisson equation for matter. However, in the D-M case $\bar{\rho}=0$ and the negative mass density $\rho_{-}$ plays the role of $\bar{\rho}$.

\section{Comoving co-ordinates}\label{sec:comoving}
Let us now consider an expanding distribution of matter with spherical symmetry.
In this case, the gravitational field has only one component $E_r(r,t)$, which depends on time and on a single spatial variable $r$. This type of system was studied extensively in the past \cite{Rouet1,Rouet2,MRexp,MR2010,miller2010ewald,Manfredi_PRE2016,quintic,Joyce2011,Benhaiem11032013}.

In the presence of a finite cosmological constant $\Lambda$, the Newtonian equation of motion for matter particles reads as:
\be
\frac{\d^2 r}{\d t^2} = E_r(r,t) +\frac{c^2\Lambda}{ 3}\,r,
\label{motion}
\ee
where $E_r=-\partial_r \phi$ is the gravitational field and $c$ is the speed of light. The factor 3 comes from the 3 space dimensions.

We consider an expanding universe with scale factor $a(t)$ and, following  \cite{Martel1998}, we define the generalized ``supercomoving" coordinates  (denoted by an overcaret) as:
\begin{eqnarray}
r &=& a(t)\, \hat r,  \label{scaling_r} \\
\d t &=& b^2(t) \,\d\hat t  \label{scaling_t}.
\end{eqnarray}
Note that we also introduced a scaled time $\hat t$, which defines a new epoch-dependent ``clock".
The velocity transforms as
\be
\frac{\d r}{\d t} = \frac{a}{b^2} \frac{\d{\hat r}}{\d\hat t} + {\dot a} {\hat r},
\label{scaling_v}
\ee
where a dot stands for differentiation with respect to $t$.
By defining $v \equiv \d r/\d t$ and $\hat v \equiv \d \hat r/\d \hat t$, Eq. \eqref{scaling_v} can be rewriten as
\be
 \frac{a}{b^2} \hat v =  v - H(t)r ,
\label{eq:vpec}
\ee
where $H(t)=\dot a/a$ is the Hubble parameter. Therefore, $v_{\rm pec}\equiv a \hat{v}/b^2$ represents the peculiar velocity, i.e., the velocity fluctuations around the Hubble flow $H(t)r$.

Using the transformations of Eqs. \eqref{scaling_r}-\eqref{scaling_t}, the comoving equation of motion becomes
\be
\frac{\d^2 {\hat r}}{\d \hat t^2}+2b^2\left(\frac{\dot a}{a}-\frac{\dot b}{b}\right)\frac{\d {\hat r}}{\d \hat t}+b^4\,\frac{\ddot a}{a}\,{\hat r}=\frac{b^4}{a^3}\, \hat{E}_r +\frac{c^2\Lambda}{ 3}\,b^4\,{\hat r}\, ,
\label{eqmotion2}
\ee
where $\hat E_r({\hat r},\hat t)$ is the scaled gravitational field. As the density must scale as $\hat\rho({\hat r},\hat t) = a^{3}(t)\rho(r,t)$ in order to preserve the total mass, we scale the gravitational field as $\hat{E}_r({\hat r},\hat t)= a^{2}(t)E_r(r,t)$, so that the Poisson equation remains invariant in the scaled variables.

We choose the same time scaling that was used for the EdS universe \cite{Manfredi2018}, i.e. $b^4=a^3$, so that the coefficient in front of the gravitational field is time-independent. This yields:
\be
\frac{\d^2 {\hat r}}{\d \hat t^2}+{1\over 2} a^{1/2}\,\dot a\, \frac{\d {\hat r}}{\d \hat t}
+a^2\,\ddot a\,{\hat r}= \hat{E}_r +\frac{c^2\Lambda}{ 3}\,a^3\,{\hat r}\, .
\label{eqmotion4}
\ee
The above comoving equation of motion can be used for both the D-M and the \LCDM cosmologies, by taking the respective scale factors $a(t)$.

\subsection{Dirac-Milne universe}\label{sec:diracmilne}
In the Dirac-Milne cosmology, the cosmological constant vanishes ($\Lambda = 0$) and the scale factor $a(t)$ is linear in time (coasting universe):
\be
a(t)= t/t_0 ,
\label{eq:aDM}
\ee
where $t_0$ denotes the present epoch. We note that, for the D-M universe, the age of the universe is {\em exactly} equal to $t_0=H_0^{-1}$, where $H_0$ is the Hubble constant at the present time \cite{Benoitlevy}.

For positive-mass particles, the equation of motion \eqref{eqmotion4} then becomes:
\be
\frac{\d^2 {\hat r}}{\d \hat t^2}+{H_0\over 2} a^{1/2}\,\frac{\d {\hat r}}{\d \hat t}= \hat{E}_r\, .
\label{eq:motion-dm}
\ee
Next, we consider a locally planar perturbation embedded in this expanding universe.
In this locally planar system, the comoving version of Poisson's equation \eqref{poiss_diracmilne3} can be approximated by its one-dimensional (1D) counterpart (we still use the notation $\hat{r}$ for the local comoving co-ordinate):
\be
\frac{\partial \hat{E}_r}{\partial \hat{r}} = -4\pi G\left [\hat\rho(\hat r, \hat t) -\rho_0 \right],
\label{eq:poisson1d}
\ee
where we recall that $\rho_0$ represents, in comoving co-ordinates, a uniform background of negative-mass particles.
The numerical results  of the D-M universe presented in the next sections will be based on N-body simulations of the scaled equation of motion \eqref{eq:motion-dm} and Poisson's equation \eqref{eq:poisson1d}.

\subsection{\LCDM universe}\label{sec:LCDM}
For the \LCDM universe, the cosmological constant is non-zero and the scale factor is a solution of the Friedmann equations (neglecting radiation):
\begin{eqnarray}
\ddot{a} &=& -H_0^2 \left({\Omega_M \over 2}\frac{1}{a^2} - \Omega_\Lambda a \right),
\label{friedmann1}\\
\frac{\dot{a}}{a} &=& H_0 \left( \frac{\Omega_M}{a^3} + \Omega_\Lambda \right)^{1/2}.
\label{friedmann2}
\end{eqnarray}
where $H_0=(\dot{a}/a)_0$.
The quantities $\Omega_M$ and $\Omega_\Lambda$ are, respectively, the densities of matter (both baryonic and dark) and vacuum normalized to the critical density, with $\Omega_{\Lambda}=\Lambda c^2/(3H_0^2)$. Approximate values \cite{tanabashi_pdg} in the \LCDM model are $\Omega_M=0.3$ and $\Omega_\Lambda=0.7$, which will be used in the forthcoming simulations.
For a case without radiation, an analytical solution of the Friedmann equation \eqref{friedmann2} can also be obtained, see Appendix \ref{app:A}.

Inserting the Friedmann equations \eqref{friedmann1}-\eqref{friedmann2} into the equation of motion \eqref{eqmotion4}, we obtain
\be
\frac{\d^2 {\hat r}}{\d \hat t^2}+
{H_0 \over 2} \left(\Omega_M + \Omega_\Lambda a^3 \right)^{1/2}\frac{\d {\hat r}}{\d \hat t} =
{\omega_{J0}^2 \over 3}{\hat r}+ \hat{E}_r\, ,
\label{eqmotion6}
\ee
where $\omega_{J0}=\sqrt{4\pi G \rho_0}$ is the Jeans' frequency and we used the following relation:
\be
H_0^2 \,\Omega_M = {2 \over 3} \,\omega_{J0}^2 .
\label{eq:omegaJ0}
\ee

For a homogeneous density, the gravitational field is $\hat{E}_r=-(\omega_{J0}^2/3)\, \hat r$, and the two terms on the right-hand side of Eq. \eqref{eqmotion6} exactly cancel each other. As in the D-M case, we assume a locally planar perturbation embedded in this expanding universe.
In this locally planar system, the factor 1/3 can be dropped from the first term on the right-hand side of Eq. \eqref{eqmotion6} and this term can be incorporated into an approximate 1D Poisson's equation
\be
\frac{\partial \hat E}{\partial \hat r} = -4\pi G \left[\hat\rho(\hat r, \hat t) - \rho_0\right].
\label{eq:poissonLCDM}
\ee
The 1D equation of motion then becomes
\be
\frac{\d^2 {\hat r}}{\d \hat t^2}+
{H_0 \over 2} \left(\Omega_M + \Omega_\Lambda a^3 \right)^{1/2}\frac{\d {\hat r}}{\d \hat t} =
 \hat{E}_r\, .
\label{eqmotionLCDM}
\ee
For $\Omega_{\Lambda}=0$ and   $\Omega_{M}=1$, we recover the EdS universe \cite{Manfredi2018}.
The \LCDM simulations presented in the forthcoming section will be performed in this planar reference frame, using Eqs. \eqref{eq:poissonLCDM} and \eqref{eqmotionLCDM}.

Interestingly, the equations of motion for the D-M universe [Eq. \eqref{eq:motion-dm}] and for  the \LCDM universe [Eq. \eqref{eqmotionLCDM}] only differ in the coefficient of the fictitious friction terms (the respective Poisson equations are also identical). This friction  coefficient can be written as a function of the scale factor (leaving aside the prefactor $H_0/2$):
\begin{eqnarray}
f_{\rm D-M} (a) &=&  a^{1/2}, \nonumber\\
f_{\rm \Lambda CDM} (a) &=&  \left(\Omega_M + \Omega_\Lambda a^3 \right)^{1/2}, \label{eq:friction}\\
f_{\rm EdS} (a) &=& 1,\nonumber
\end{eqnarray}
where we also added the EdS case for comparison.
Note that, for the D-M and \LCDM cosmologies, the  friction coefficient grows with $a$, which will lead to the freezing of the gravitational structures before or around the present epoch (roughly, $z \approx 3$ for D-M and $z \approx 0.5$ for $\Lambda$CDM).

\begin{figure}[]
 \begin{center}
\includegraphics[width=0.6\linewidth]{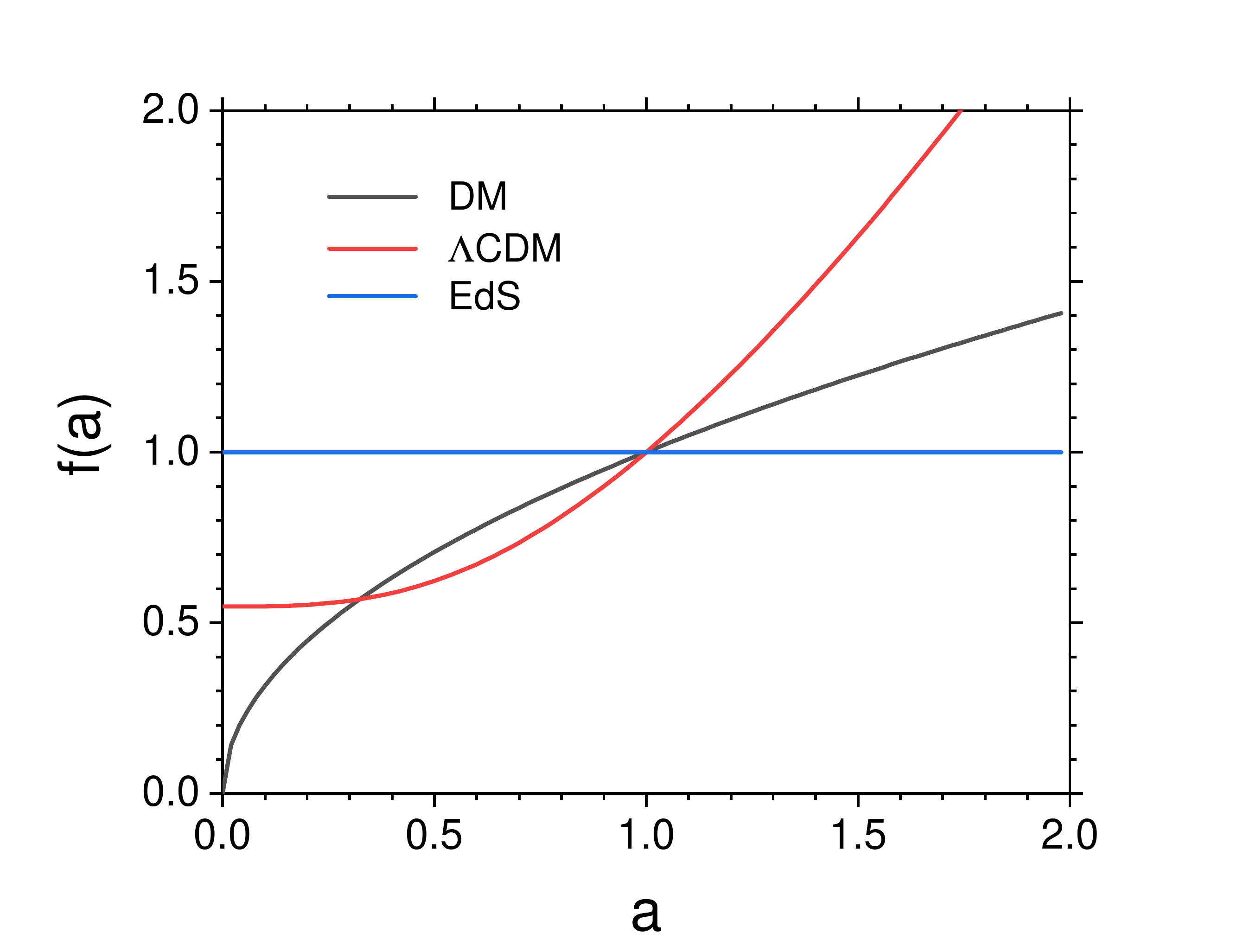}
 \end{center}
    \caption{Coefficient $f(a)$ of the friction term in the various equations of motion, see Eq. \eqref{eq:friction}.} \label{fig:fa}
\end{figure}

The three cases are represented in Fig.~\ref{fig:fa}. The friction term is weaker at early times ($a \lesssim 0.3$) for the D-M model compared to \LCDM leading to a faster formation of larger structures in the D-M universe, as will be shown in the forthcoming simulations. For $0.3 \lesssim a < 1$, the situation is reversed.
Interestingly, the integrated coefficient between the Big Bang and the present time
\[
\bar f = \int_0^1 f(a) \,da,
\]
is almost identical for the D-M universe ($\bar f  = 0.666$) and the \LCDM universe ($\bar f = 0.676$).
This is an interesting indication that, although the two models have very different past histories, they should lead to a similar universe at the present time. In particular, they should both stop forming structures at similar epochs, as will be confirmed by the forthcoming numerical simulations.

\section{Simulation results}\label{sec:numres}
In this section we present the results of numerical simulations obtained with the Dirac-Milne and \LCDM models described in the preceding sections. The simulations were performed with an N-body code \cite{Manfredi2018} that solves the relevant equations of motion \eqref{eq:motion-dm}  or  \eqref{eqmotionLCDM} for $N$ interacting particles, using a velocity Verlet scheme. Typical simulations employed $N \approx 2.5\times 10^5$ particles.
To reduce the level of fluctuations, for each case we performed ensemble averaging over 5 statistically equivalent initial conditions.

As mentioned in the preceding sections, we consider a 3D expanding spherically-symmetric universe and then study planar perturbations in the comoving coordinates. This  reduces the problem to one spatial dimension in the local comoving coordinate $\hat r$, which will be represented in the numerical results. In this 1D approximation, the particles are in fact infinite sheets with uniform surface mass density. Boundary conditions are taken to be spatially periodic.
More details on the model can be found in Refs. \cite{MRexp,MR2010,miller2010ewald,Manfredi_PRE2016,Manfredi2018}.
As already stated (see Sec. \ref{sec:Modeldiracmilne}), only matter (positive mass) is evolved in the numerical simulations. Antimatter (negative mass) is approximated by a uniform background with constant density $\rho_0$ in the comoving frame. In the present set of simulations, no attempt is made to simulate the depletion zone between matter and antimatter.

For both models, the initial condition is set at recombination, corresponding to redshift $z=1080$; note that this redshift does not correspond to the same cosmological time in the D-M ($\approx 14$ million years) and \LCDM ($\approx 380\,000$ years) universes, see \cite{Manfredi2018}.
The initial matter density is the sum of a spatially uniform term $\rho_0$ and a small perturbation $\tilde\rho$ with the 1D power spectrum $P_{\rm 1D}(k) = |\tilde\rho_k|^2 \sim k^p$, where $k$ is the wavenumber. Initial power spectra of this form, with $p \in [0,4]$, were used in a number of earlier works on structure formation \cite{Joyce2011,MR2010}.
In the present work, we take $p=3$, which produces a spectrum that is largest at small wavelengths, and then study the clustering of matter at increasingly larger scales.
The standard 3D power spectrum $P(k)$ can be obtained by setting: $P_{\rm 1D}(k)dk = P(k) \,2\pi  k^2 dk$, yielding $P(k) = P_{\rm 1D}/(2\pi k^2)$. Hence, the initial  3D spectrum behaves as  the standard Harrison-Zeldovich spectrum:  $P(k)\sim k$.
The decrease of $P(k)$ at $z=0$ for large wavenumbers originates from the fact that the initial 1D spectrum becomes flat due to statistical noise when $2\pi/k$ reaches the average initial interparticle distance.

For the sake of comparison, the initial spectra are taken to be statistically identical (same slope)  for the D-M and \LCDM models. The ensuing evolution is fully nonlinear in both cases. This is the expected behavior in the D-M universe, where the density fluctuations are already large at $z=1080$ \cite{Benoitlevy}, so that collapse
and nonlinear evolution occur almost immediately.
In contrast, in the standard cosmological model, the evolution of the power spectrum is almost entirely linear, the nonlinearity bringing only a relatively minor correction \cite{Aubourg2015}.
Here, however, the idea is to closely compare the two universes starting from the same initial configuration. Therefore, we choose to focus only on the nonlinear development of structures even for the $\Lambda$CDM case.

The initial velocities are set using the Zeldovitch approximation, so that only the growing mode is excited \cite{MR2010}.
The influence of the initial condition for the particle velocities has been studied by varying the amplitude of the velocity distributions.  For the Dirac-Milne model, the uncertainty and influence of these initial velocities should be relatively small, as the initial structures are initiating their nonlinear evolution almost immediately after the CMB transition, which results in rapid virialization of the velocities of the initial structures, effectively decoupled from the cosmological expansion.
The initial velocity spread for the simulations presented here can be read from the corresponding figure at the end of this section (Fig.~\ref{fig:vthpeculiar}).

The evolution is labeled by either the scale factor $a$ or the cosmological redshift $z=(1-a)/a$.
In the numerical code, densities are measured in units of $\rho_0$ and time in units of $\omega_{J0}^{-1}$.
Lengths are measured in units of an arbitrary length scale $\lambda$ and  the gravitational field (an acceleration) is expressed in units of $\lambda\omega_{J0}^{2}$. In practice, $\lambda$ is an adjustable parameter which is chosen here so that the power spectrum at $z=0$ issued from the simulations has a peak at the same wavelength as the spectrum obtained from observations, such as the Sloan Digital Sky Survey (SDSS) \cite{Tegmark2004}.
Then, once $\lambda$ has been fixed, all other normalizations (e.g., for the particle velocities or the amplitude of the power spectrum) follow uniquely without any additional assumptions.

We show the results of two typical  simulations, one each  for the D-M and  \LCDM models.

\begin{figure}[]
 \begin{center}
\includegraphics[width=0.7\linewidth]{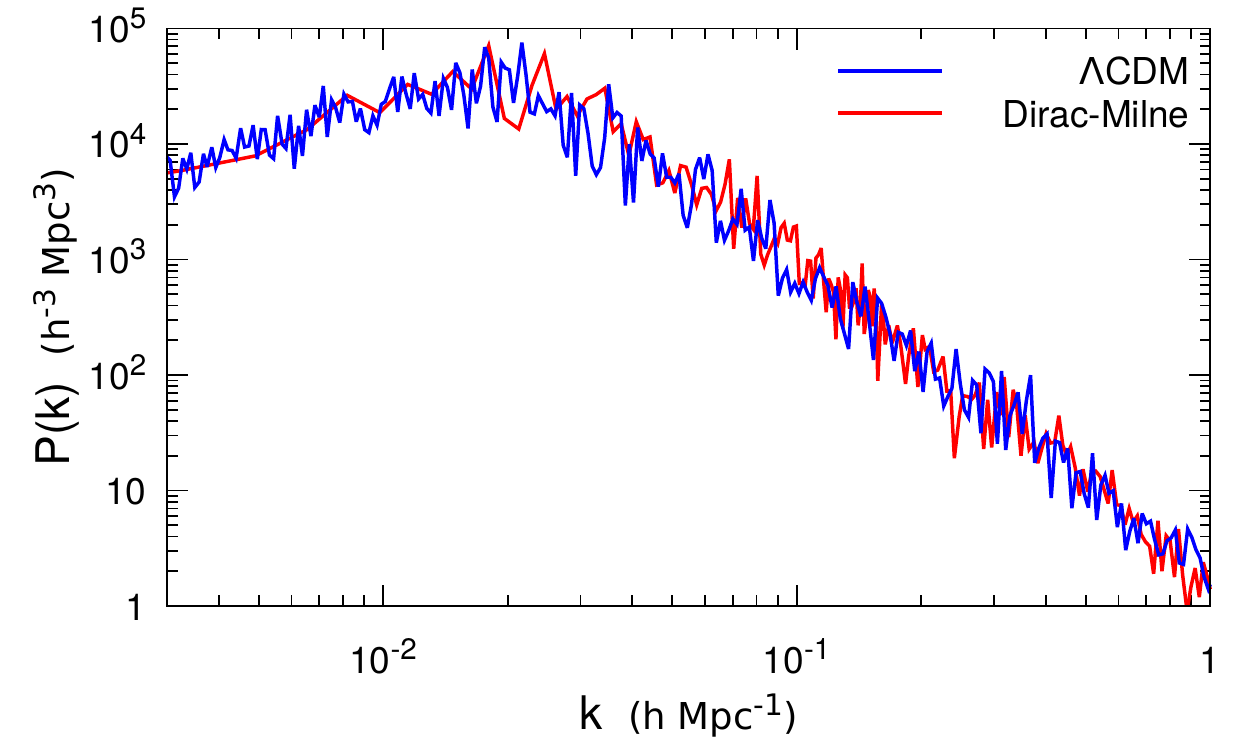}
 \end{center}
    \caption{Power spectra at $z=0$ for the D-M and \LCDM universes.} \label{fig:twospectra}
\end{figure}

The power spectra at $z=0$ for  the two cosmologies are shown synoptically in Fig.~\ref{fig:twospectra}. As mentioned above, the horizontal axis has been scaled so that the spectra peak at $0.018 h\rm\, Mpc^{-1}$, where $h=H_0/(100 \,\rm km/ s/Mpc)$, as in the observed SDSS spectrum \cite{Tegmark2004}.
The shape of the contemporary D-M and \LCDM spectra are virtually identical, supporting the idea that, although the details of the evolutions are very different (coasting expansion for D-M as opposed to a sequence of accelerations and decelerations for \LCDM), the end result at $z=0$ is rather similar. This conclusion is in line with our earlier observation (see Fig.~\ref{fig:fa}) that, while the respective comoving equations of motions are different, ``averaging" (in  a loose way) between $a=0$ and $a=1$ produces effectively the same results.
Further, the expected slopes for long ($P \sim k$) and short ($P \sim k^{-3}$) wavelengths are correctly recovered by the Dirac-Milne simulations.
This power law behavior in the nonlinear regime suggests a self-similar matter distribution in each model, pointing to the existence of a robust fractal dimension \cite{MR2010,Shiozawa2016}.

\begin{figure}[]
 \begin{center}
\includegraphics[width=0.7\linewidth]{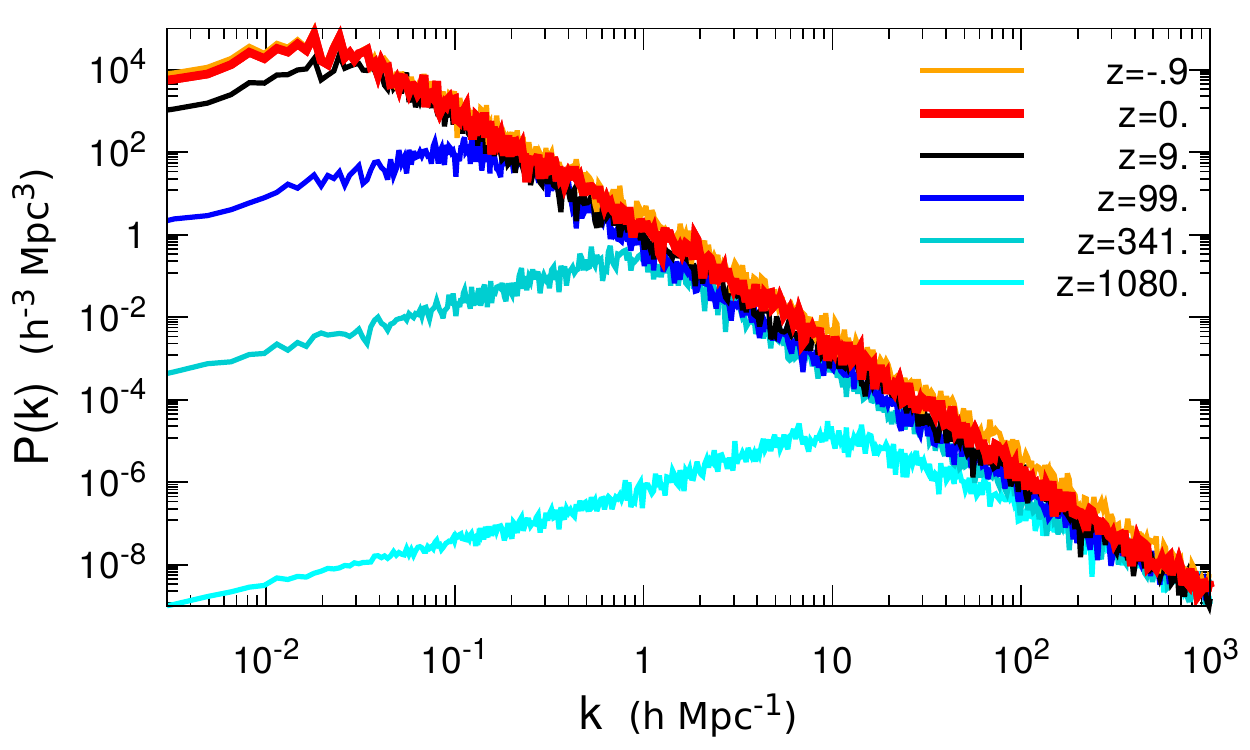}
\includegraphics[width=0.7\linewidth]{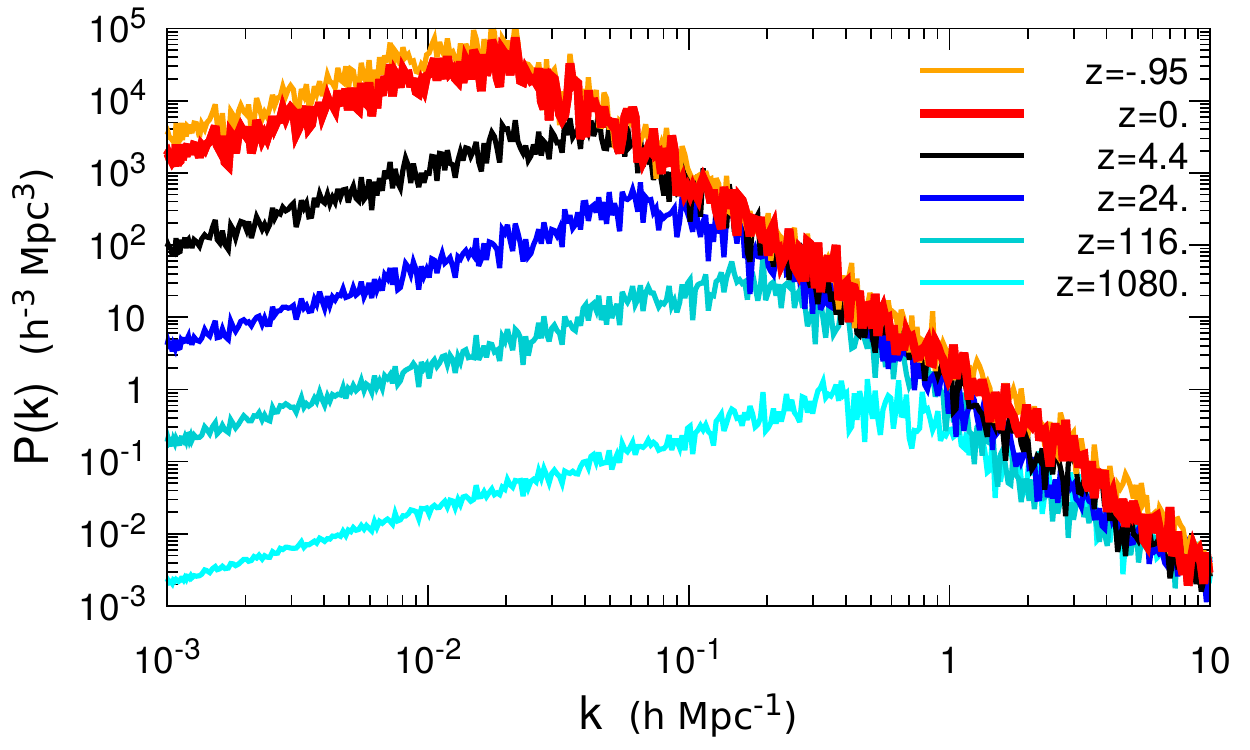}
 \end{center}
    \caption{Evolution of the power spectra for the cases D-M (top panel) and $\Lambda$CDM (bottom panel), for different cosmological redshifts $z$. The thick lines correspond to the present epoch ($z=0$).
    A negative value of $z = -0.9$ for D-M corresponds to $a = 10$ ($t\approx 140\, \rm Gy$), while a negative value of $z = -0.95$ for \LCDM corresponds to $a = 20$ ($t\approx 65\, \rm Gy$).} \label{fig:spectraevol}
\end{figure}

The evolution of the power spectra  from $z=1080$ to $z \approx -0.9$ (corresponding to $a \approx 10$) is displayed in Fig.~\ref{fig:spectraevol}. For both cosmologies, it is clear that structure formation has stopped at, or slightly earlier than, the present epoch.
For the D-M case (Fig.~\ref{fig:spectraevol}, top panel), the initial spectrum at $z=1080$ peaks around $100\, h^{-1} \rm kpc$, when structure formation begins. Then, the evolution proceeds by collecting larger and larger clusters in a bottom-up fashion.
At the present epoch, the  spectrum displays a nonlinear power-law behavior extending over four decades in wave number space, with a peak around $50\, h^{-1} \rm Mpc$, the formation of the largest structures being frozen since $z \approx 3$ in the D-M universe.
We stress again that the transition between the  $P \sim k$ and the $P \sim k^{-3}$ regimes comes about because of strong nonlinear effects -- whereby clusters of matter coalesce into bigger clusters,  then into even bigger clusters, and so on and so forth -- until this process stops shortly before the present epoch. For D-M, this freezing of structures is due to the presence of a homogeneous background of negative mass, which acts as a cosmological constant that decreases with time, as was mentioned in Sec. \ref{sec:Modeldiracmilne}.

A similar nonlinear build-up of gravitational structures is also seen in our \LCDM simulations (Fig.~ \ref{fig:spectraevol}, bottom panel), in contrast with the standard \LCDM model which is essentially linear \cite{Aubourg2015}. This is due to our choice of initial condition, dictated by our wish to highlight the differences between D-M and \LCDM starting from a similar configuration.

\begin{figure}[]
 \begin{center}
\includegraphics[width=0.65\linewidth]{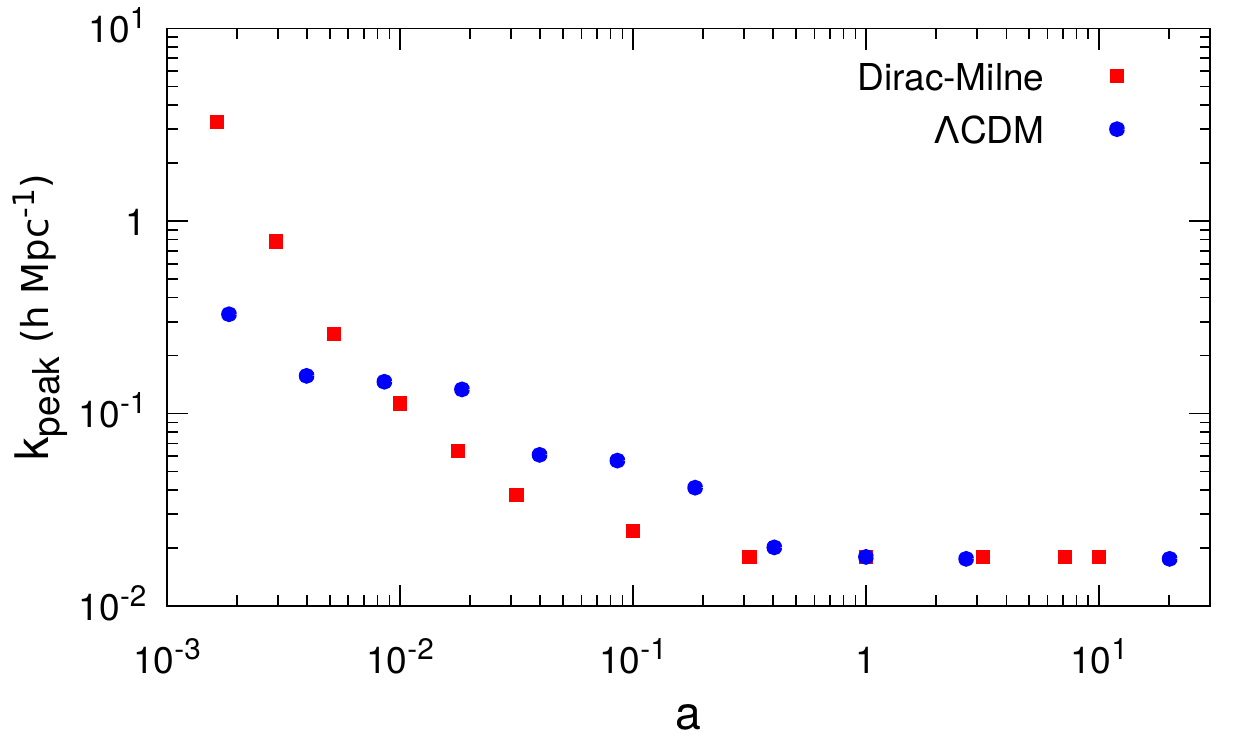}
\includegraphics[width=0.65\linewidth]{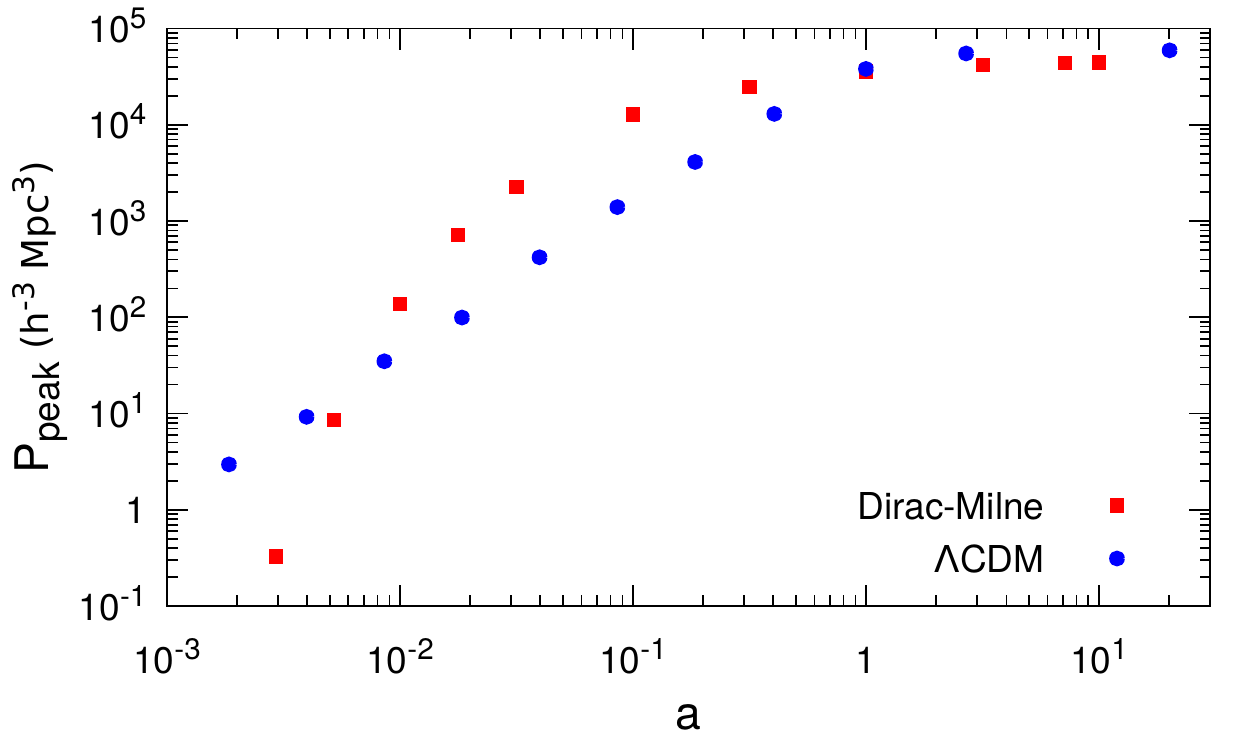}
 \end{center}
    \caption{Top panel: Wave number $k_{\rm peak}$ corresponding to the peak of the power spectra for the Dirac-Milne and \LCDM universes as a function of the scale factor $a(t)$. {\gm Bottom panel: Corresponding  value of the peak power at $k_{\rm peak}$}. A striking feature of these simulations is that $k_{\text{peak}}$ evolves in time, describing the nonlinear evolution in both the D-M and \LCDM models, whereas in the standard \LCDM analysis the usual assumption is that the nonlinear evolution represents only a small correction, while the peak position  is fixed at $k_{\text{peak}}\sim 0.018\,h\text{Mpc}^{-1}$ corresponding to the mode entering the horizon at matter-radiation equality.} \label{fig:kpeak}
\end{figure}

As was already noticed in our earlier work \cite{Manfredi2018}, structure formation is  initially faster for the D-M universe.
This fact is even more apparent from the evolution of the peak wave number of the power spectrum, as shown in Fig.~\ref{fig:kpeak}, where we clearly see that structure formation has stopped at a similar epoch for both universes. {\gm The peak power also grows faster in the D-M universe, and is larger than the corresponding \LCDM power during the epochs between $z \approx 100$ and $z=0$. This discrepancy might yield a difference in the predicted abundance of clusters or quasars at high redshift, which could be tested against observation.}
Again, it is important to note that the evolution in time of the position of the peak for our \LCDM simulation reflects the development of structures in the nonlinear regime,  while in the standard \LCDM analysis, the peak position is time-independent and fixed by the mode entering the horizon at matter-radiation equality \cite{Aubourg2015}.

\begin{figure}[]
 \begin{center}
\includegraphics[width=\linewidth]{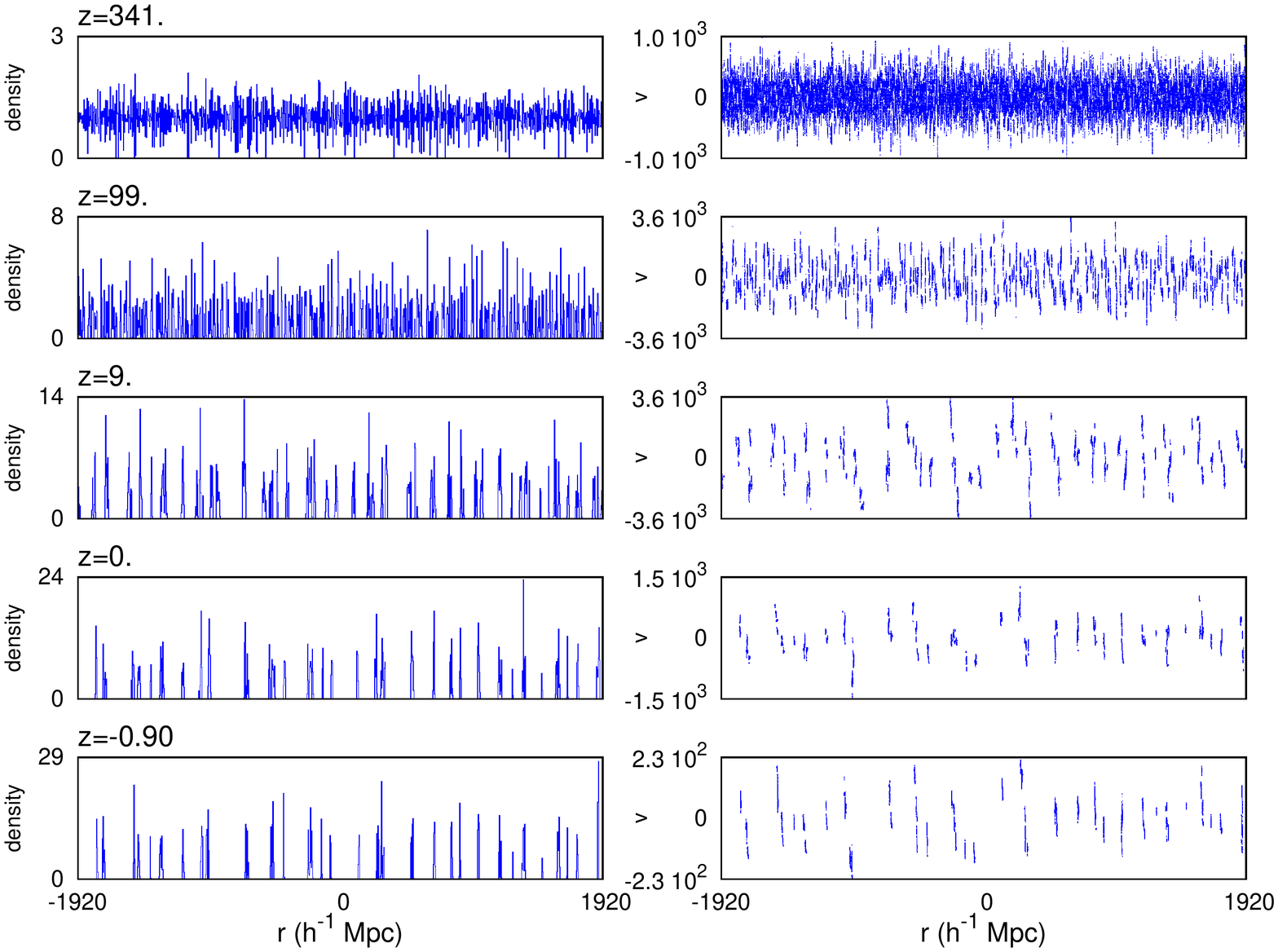}
 \end{center}
    \caption{Dirac-Milne universe. Left frames: Matter density $\rho(\hat{r},t)$ in comoving space for the positive-mass particles, at five epochs characterized by different redshift $z$. Right frames: Corresponding distributions in the phase-space. The vertical axis is the peculiar velocity, in km/s.} \label{fig:phasespaceDM2}
\end{figure}

\begin{figure}[]
 \begin{center}
\includegraphics[width=\linewidth]{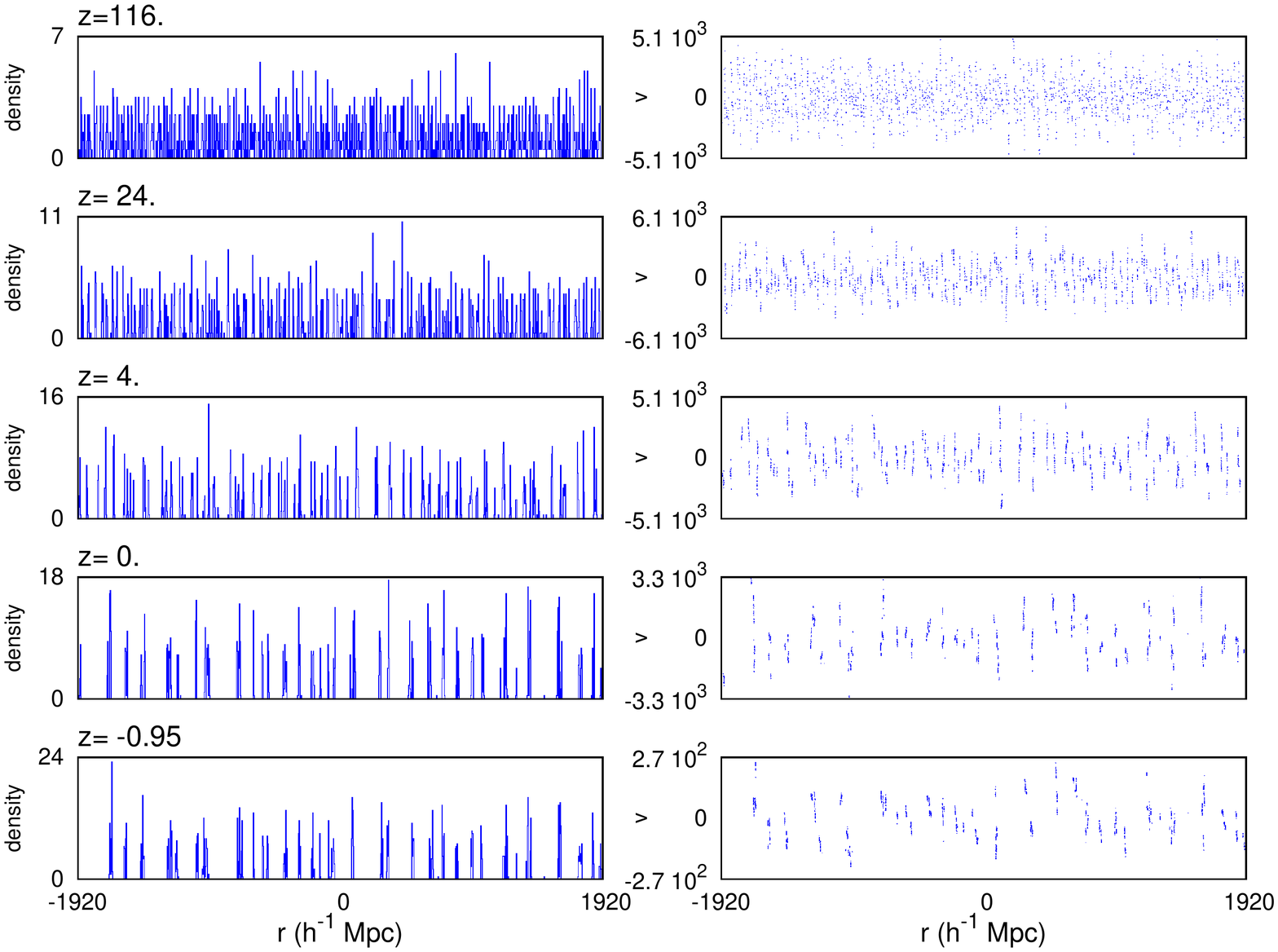}
 \end{center}
    \caption{\LCDM universe. Left frames: Matter density $\rho(\hat{r},t)$ in comoving space for the positive-mass particles, at five epochs characterized by different redshift $z$. Right frames: Corresponding distributions in the phase-space. The vertical axis is the peculiar velocity, in km/s.} \label{fig:phasespaceLCDM2}
\end{figure}

\begin{figure}[]
 \begin{center}
\includegraphics[width=0.65\linewidth]{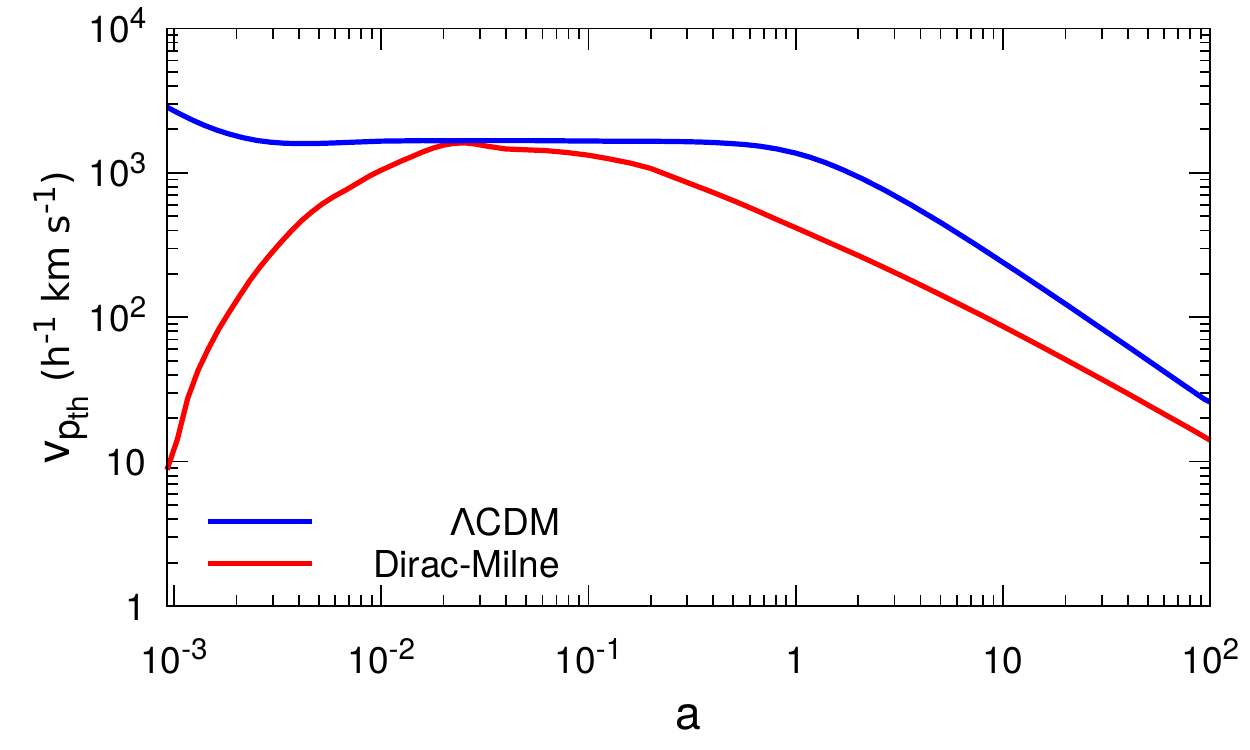}
 \end{center}
    \caption{Dispersions of the peculiar velocities for the D-M and  nonlinear \LCDM cosmologies,
   as a function of the scale parameter $a(t)$.} \label{fig:vthpeculiar}
\end{figure}

Structure formation in comoving space is shown in Figs. \ref{fig:phasespaceDM2} and \ref{fig:phasespaceLCDM2} for the D-M and $\Lambda$CDM universes, respectively. Again, in both cases the gravitational structures freeze, around $z \approx 0.5$ ($a\approx  0.67$) for $\Lambda\rm{CDM}$ and $z \approx 3$ ($a \approx  0.25$) for D-M. The typical size of the contemporary structures is around 50~Mpc, in accordance with the power spectra of Fig.~\ref{fig:twospectra}. At the same epoch, the peculiar velocities start decreasing, signalling a local cooling.

Figure \ref{fig:vthpeculiar} shows the ``thermal" dispersion of the peculiar velocities, see  Eq. (\ref{eq:vpec}), defined as $v_{\rm p,th} \equiv \langle (v-Hr)^2\rangle^{1/2}$, where the average is taken over all the particles. At the present epoch ($a=1$), D-M predicts peculiar velocities of the order of 400~km/s, which is comparable with the  values usually reported in the literature \cite{Kashlinsky2008,Girardi1993}. In future epochs, significant cooling can be observed.

\begin{figure}[]
 \begin{center}
\includegraphics[width=0.65\linewidth]{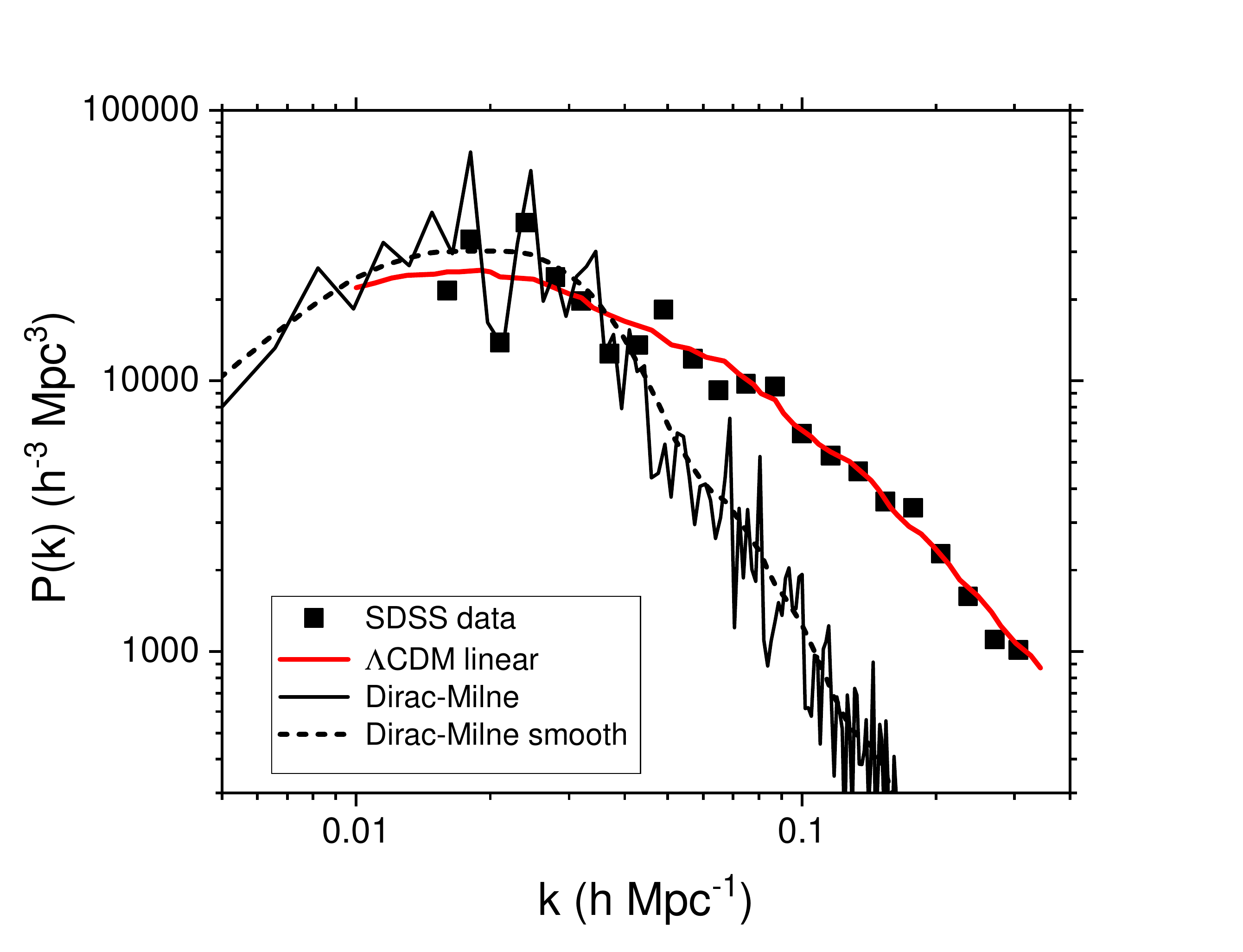}
 \end{center}
    \caption{Comparison of power spectra at $z=0$. Black squares: SDSS data points; red line: best fit linear \LCDM model {\gm (from Ref. \cite{Tegmark2004})}. Solid black line: Dirac-Milne spectrum (this work); black dashed line: Dirac-Milne spectrum averaged with a  moving triangular window over 10 points.} \label{fig:SDSS}
\end{figure}

Finally, in Fig.~\ref{fig:SDSS}, we compare the power spectrum obtained from our own Dirac-Milne simulation (solid and dashed black curves) with data points from the SDSS survey (black squares), as well as a best-fit linear \LCDM model to these data (red line), {\gm both taken from Ref. \cite{Tegmark2004}}.
{\gm Although the D-M simulated spectrum agrees relatively well with the data points around the main peak}, at shorter wavelengths it displays significantly  less power (about one order of magnitude) than the SDSS data
{\gm
(note that, as is apparent from Fig. \ref{fig:twospectra}, our nonlinear \LCDM spectrum at $z=0$ is very similar to the corresponding D-M one).
This is an important issue that will have to be addressed in future, more sophisticated, simulations. The most obvious limitations of the present approach are the 1D nature of the modeling and the absence of baryonic physics and related feedback, which are fundamental at smaller scales but probably less important for the position of the main peak at smaller $k$ (large scales).
}

\section{Conclusions and perspectives}\label{sec:conclusion}
In this paper, we compared nonlinear structure formation in the \LCDM and the Dirac-Milne cosmologies. This is a follow-up to our previous study comparing the Dirac-Milne and Einstein-de Sitter  universes, where it was shown that the structure formation in the Einstein-de Sitter universe never fully stops, this universe being critical and right at the border of recollapse. The present study highlights the similarities in the recent ($z \lesssim 5$) and contemporary stages of the D-M and $\Lambda$CDM universes: both universes stop their construction of larger structures at about the same time -- a few billion years after the Big-Bang -- and in the case of $\Lambda$CDM approximately at the epoch where the Dark Energy component is supposed to become predominant.
The gravitational structures observed around the present epoch in our simulations are surprisingly  similar for the two models. This is all the more remarkable, as the two universes have undergone a very different earlier history -- coasting expansion for D-M vs. a sequence of accelerations and decelerations in \LCDM.

{\gm
Further, we have compared the D-M power spectrum with observational data from the SDSS survey \cite{Tegmark2004}. Although the D-M spectrum agrees relatively well with the data points around the main peak, it displays significantly  less power at shorter wavelengths. This discrepancy may be attributed to the 1D geometry and lack of baryons physics and feedback effects in our approach. More complete 3D simulations, currently underway, should be able to settle this important issue.
}

It is important to note that the mechanisms of construction of large-scale structures, although similar in their contemporary stages, are remarkably different in their early stages. The Dirac-Milne cosmology begins  its structure formation very early (a few tens of millions of years after the CMB transition) and from relatively small matter pools (typically 10$^5$ solar masses, with domain size $\approx 10\rm\, pc$ comoving at $z = 1080$) \cite{Benoitlevy}, directly in a nonlinear regime with a matter-antimatter density contrast of order unity.
Then, structure formation proceeds as a bottom-up nonlinear construction, with small clusters coalescing into larger and larger ones, until this process stops near the present epoch.

In contrast, \LCDM starts its structure formation from a nearly homogeneous universe with a linear growing mode that requires an elusive Dark Matter component to trigger the onset of nonlinear collapse and reionization at a much later stage.
In this respect, Dirac-Milne appears to provide a much more unified approach, starting from a single scale and building up in a nearly pure bottom-up construction, with three orders of magnitude of comoving scale growth, the entire matter power spectrum.

{\gm
Further, the Dirac-Milne cosmology may provide an explanation for what appears in \LCDM as a remarkable coincidence, since the small oscillation of the BAOs (a few percent) occurs at a scale where strong clustering and inhomogeneity is clearly visible at the same 100-150 Mpc scale. In Dirac-Milne, there is a single scale, resulting from the inhomogeneities of the matter-antimatter emulsion at $z \approx 1080$, followed by roughly three orders of magnitude of bottom-up nonlinear growth of this initial scale.
}

Encouraged by the present study, we will present in a forthcoming publication the results of a set of 3D simulations using a modified version of the \texttt{RAMSES} simulation code \cite{Teyssier_2002}. There, we shall investigate the impact of the asymmetry between the geometrical distributions of the matter component, collapsed in planes, filaments and cluster node structures, and the antimatter clouds, unable to collapse due to their internal repulsion, and separated from matter by a depletion zone.
These further 3D simulations may allow us to investigate the origin of the baryonic acoustic oscillations (BAOs), an issue not addressed here, which  in the Dirac-Milne model have a much larger value ($\approx$ 20 Gpc) and play a completely negligible role in the formation of structures. Indeed, in the Dirac-Milne universe, all the structures develop from a single, nonlinear, self-similar process.

\noindent{\bf \large Acknowledgments}\\
We are indebted to Jim Rich for his thorough reading of the manuscript and many insightful comments. We also thank Cl\'ement Stahl for several useful suggestions. Needless to say, the authors are solely responsible for the
errors or imprecisions that may still remain in this paper.
The numerical simulations were performed on the computer cluster at the Centre de Calcul Scientifique en r\'egion Centre-Val de Loire (CCSC).

\bibliography{diracmilne_biblio}

\newpage

\appendix
\section{Integration of the Friedmann equation}\label{app:A}

For the D-M universe, the behavior of the scale factor $a(t)$ is known analytically, see Eq. \eqref{eq:aDM}.
For the \LCDM case, one has to solve the Friedmann equation (neglecting radiation, as was done throughout this work) in order to obtain $a(t)$.
As our equation of motion \eqref{eqmotionLCDM} is written in comoving co-ordinates, it is  convenient to rewrite the Friedmann equation  \eqref{friedmann2} using comoving variables. With the help of  Eq. \eqref{scaling_t}, we get:
\be
\frac{\d a}{\d \hat t}=H_0\, a\, \sqrt{\Omega_M+\Omega_{\Lambda}a^3}.
\label{eq.an.0}
\ee
Integrating Eq. \eqref{eq.an.0} yields
\be
\int_{a_i}^a \frac{1}{a'\sqrt{1+\frac{\Omega_{\Lambda}}{\Omega_M}a'^3}} \d a'=\sqrt{\frac{2}{3}}\,\omega_{J0}
\int_0^{\hat t} \d \hat t' ,
\ee
where $a_i=1/(z_i+1)$, with $z_i=1080$, is the initial value of the scale factor and we have introduced the Jeans frequency $\omega_{J0}$ using Eq. \eqref{eq:omegaJ0}. We obtain:
\be
\log\frac{\sqrt{1+\Omega_{\Lambda}a(\hat t)^3/\Omega_M}-1}{\sqrt{1+\Omega_{\Lambda}a(\hat t)^3/\Omega_M}+1}-\log\frac{\sqrt{1+\Omega_{\Lambda}a_i^3/\Omega_M}-1}{\sqrt{1+\Omega_{\Lambda}a_i^3/\Omega_M}+1}=\sqrt{6}\, \omega_{J0}\,\hat t,
\label{eq.an.1}
\ee
which gives
\be
a(\hat t)=\left[\frac{4\Omega_M}{\Omega_{\Lambda}}\frac{g_i\exp(\sqrt{6}\, \omega_{J0}\,\hat t)}{\left(1-g_i\exp(\sqrt{6}\, \omega_{J0}\,\hat t)\right)^2}\right]^{1/3},
\label{eq.an.2}
\ee
with
\be
g_i=\frac{\sqrt{1+\Omega_{\Lambda}a_i^3/\Omega_M}-1}{\sqrt{1+\Omega_{\Lambda}a_i^3/\Omega_M}+1}.
\label{eq.an.3}
\ee
Note that Eq. \eqref{eq.an.1} has a vertical asymptote: when $a \to \infty$, then $t \to t_{\infty}$, where
\[
\omega_{J0}\,\hat t_{\infty}=-\frac{1}{\sqrt{6}}\, \log(g_i)
\]
With the parameters of the present work, $\omega_{J0} \hat t_{\infty} \approx 8.775$.

Once the expression of $a(\hat t)$ is known, it is possible to integrate Eq. \eqref{scaling_t} to obtain the relation between the real time $t$ and the scaled time $\hat t$:
\be
H_0 t=\frac{2}{3\sqrt{\Omega_\Lambda}}\log\left(\frac{1+\sqrt{g_i}\exp(\sqrt{3/2}\, \omega_{J0}\,\hat t)}{1-\sqrt{g_i}\exp(\sqrt{3/2}\, \omega_{J0}\,\hat t)}\right).
\label{eq.an.4}
\ee
For $\hat t=0$, Eq. \eqref{eq.an.4} gives the initial time $t_i$, corresponding to the recombination epoch.

\begin{figure}[]
 \begin{center}
\includegraphics[width=0.7\linewidth]{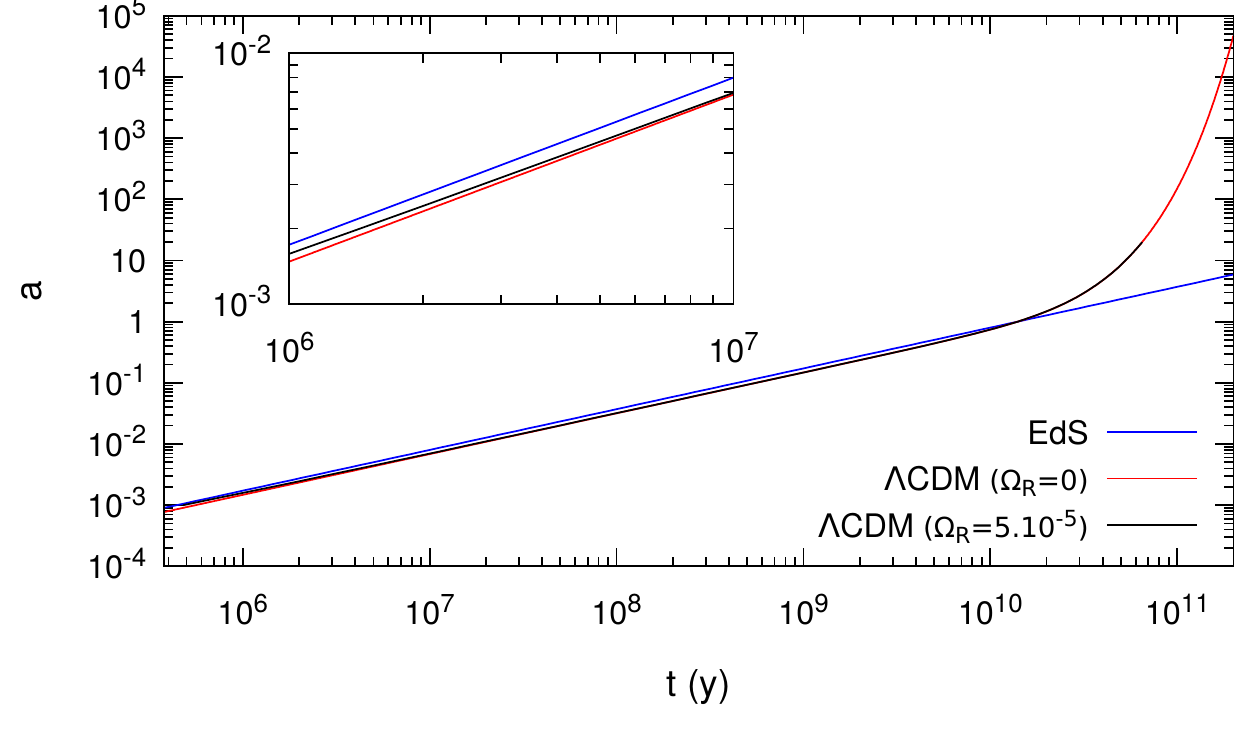}
 \end{center}
    \caption{Time evolution of the scale factor $a(t)$ for the EdS universe and for the \LCDM universe without ($\Omega_R =0$) and with ($\Omega_R = 5 \times 10^{-5}$) radiation. The inset is a zoom at early epochs.} \label{fig:friedmann}
\end{figure}

If one includes radiation in the model (i.e., $\Omega_R >0$), no analytical solution can be found, and the Friedmann equation should be integrated numerically. However, the difference is minimal, as is shown in Fig.~\ref{fig:friedmann}, which displays the evolution of the scale factor for $\Omega_R =0$ and $\Omega_R =5 \times 10^{-5}$, together with the expression for an EdS universe ($a \sim t^{2/3}$).

\end{document}